\documentclass[runningheads]{llncs}
\usepackage[T1]{fontenc}
\usepackage{xcolor}
\usepackage[colorlinks=true, urlcolor=blue, linkcolor=blue, citecolor=blue]{hyperref}
\usepackage{booktabs}
\usepackage{multirow}
\usepackage{array}
\usepackage{makecell}
\usepackage{float}
\usepackage{subcaption}

\usepackage{graphicx}
\usepackage[most]{tcolorbox}

\newtcolorbox{pracnote}{
    colback=gray!5!white,
    colframe=gray!75!black,
    title=Practitioner Notes,
    fonttitle=\bfseries,
    arc=0mm,
    breakable
}

\begin{document}
\title{Exploring How Agent Voice Accents Shape Human-AI Collaboration in K-12 Group Learning}
\titlerunning{How Agent Voice Accents Shape K-12 Human-AI Collaboration}
%
%
\author{Prerna Ravi, Carúmey Stevens, Ben Hurt, Brandon Hanks, Grace Lin, and Emma Anderson} 
%
\authorrunning{Ravi et al.}
%
\institute{Massachusetts Institute of Technology, Cambridge MA 02139, USA}
%
\maketitle              

\begin{abstract}
Collaboration is widely recognized as a cornerstone of 21st-century education, yet teachers still encounter persistent challenges in fostering productive peer interaction.
LLM conversational peer agents introduce new possibilities for mediating in-person group work, raising questions about how persona design—particularly their voice characteristics—shapes learners’ perceptions, trust, and interactional dynamics. While prior work has examined agent accent effects in one-to-one settings, little is known about how these effects manifest in groups.
We conducted a between-subjects mixed-methods study with 33 teachers examining how a GenAI voice agent with different accents (British, Indian, and African American) influenced collaboration and agent perception.
Across surveys, group interaction analyses, and artifacts, we find that accent shaped participants’ mental models and the roles the agent assumed in group interaction. The British-accented agent was largely treated as a tool and engaged in detached, utility-based ways, whereas Indian- and African American-accented agents were more readily anthropomorphized and integrated as peers. These role expectations influenced trust, engagement, and reliance over time.
This work advances understanding of how GenAI’s sociolinguistic design features shape group dynamics in CSCL, with implications for designing culturally inclusive AI partners in group learning.

\end{abstract}

\begin{pracnote}
    What is already known about this topic

    \begin{itemize}
        \item Collaborative learning supports communication, critical thinking, and shared problem-solving, but teachers often struggle with uneven participation and managing group dynamics.
        \item Conversational AI peer agents can scaffold discussion, prompt reflection, and provide feedback in classroom settings.
        \item Voice characteristics such as accent influence how users perceive agent credibility, warmth, and competence in one-to-one human–AI interactions.
    \end{itemize}

    What this paper adds

    \begin{itemize}
        \item In group settings, an AI agent’s voice accent shapes how teachers conceptualize its role (e.g., tool, peer, authority), influencing trust and engagement.
        \item Early classroom activities strongly anchor how teachers position an AI partner, affecting whether it is integrated into collaborative dialogue.
        \item Trust in group-facing AI depends not only on accuracy and responsiveness, but on alignment between the agent’s perceived role, behavior, and sociolinguistic cues.
    \end{itemize}

    Implications for practice and/or policy

    \begin{itemize}
        \item When introducing AI partners in classrooms, explicitly frame the agent’s role and set norms for how it should participate in group work. 
        \item Consider agent voice design (accent and modality) as a pedagogical choice, as it may shape authority, inclusion, and participation patterns in groups.
        \item Design introductory activities that encourage collaborative dialogue rather than system testing, helping the AI agent function as a learning scaffold rather than a background tool.
    \end{itemize}
    
\end{pracnote}

\section{Introduction}

Collaboration is central to 21st-century education, supporting communication, critical thinking, and socio-emotional development \cite{liu2024}. However, supporting classroom collaboration remains challenging for teachers due to uneven participation, group conflict, and difficulties of fairly assessing individual contributions \cite{Roberts2007}. The rise of generative AI (GenAI) introduces new possibilities for scaffolding collaborative learning. Large language models (LLMs) can provide context-sensitive, dialogic feedback \cite{lu2025optimizing} and are increasingly positioned in classrooms not only as assistants but as active collaborators in group learning \cite{Zhang2025}.
However, important gaps remain. 

First, few studies examine LLM conversational agents (CAs) designed as \textit{near-peers} in group learning rather than tutors or tools. Although peer agents can foster equitable participation, creativity, and knowledge-sharing \cite{lyu2026designing,perez2021review}, little is known about how their persona influences group dynamics \cite{liu2024}. Prior research on AI peers has largely centered on students, with limited attention to teachers’ perspectives, even though teachers ultimately mediate classroom adoption and pedagogical framing of such systems.

Second, while voice characteristics are central to AI CA persona \cite{holliday2023siri,pal2019user}, their role in group contexts is underexplored. Specifically, agent accents influence perceptions of warmth, competence, and trust in one-on-one interactions \cite{pycha2024influence,krenn2017speak}, yet little is known about how these effects unfold in multi-party collaboration, where authority, trust, and peer dynamics are collectively negotiated \cite{Zhang2023Investigating}.

We conducted a between-subjects mixed-methods study asking: \textbf{RQ- How do different voice accents of AI CAs shape perceptions and group dynamics in collaborative settings?} 
We deployed a GenAI-powered audio agent with varying accents to 33 K-12 teachers working on small-group problem-solving tasks. Accents significantly shaped teachers’ mental models of the agent—as a tool, expert, or peer—which influenced engagement and trust in group decision-making. We highlight sociolinguistic design as a critical yet understudied factor in AI-supported classroom collaboration.

\section{Related Work}
\subsection{AI-Supported Collaborative Learning}

Collaboration involves two or more students working interdependently toward a shared goal, pooling their knowledge and efforts to achieve outcomes beyond individual work \cite{evans2020measuring}. 
It fosters adaptive soft skills such as teamwork, communication, negotiation, and socio-emotional development \cite{hernandez2019computer}, yet remains difficult to implement due to free-riding, skill disparities, group composition challenges, and fair assessment concerns \cite{Roberts2007}.


LLMs enable more fluid dialogue \cite{liu2024}, supporting richer group discussion \cite{anderson2025exploring}, timely peer prompting \cite{de2025investigating}, and collective decision-making \cite{Chiang2024EnhancingAI}. Conversational agents are increasingly being positioned as team members rather than passive tools \cite{Amiot2025,Zhang2025}.
\textit{Peer} agents in groups can foster emotional regulation, equitable participation, and challenge learners’ thinking \cite{liu2024,perez2021review}, but also raise concerns around cognitive dependency, autonomy, bias, and creativity \cite{Zhang2025}.
Despite growing work with students on peer CAs, teachers who shape technology integration remain underexplored \cite{zawacki2019systematic}. We examine teachers’ perspectives on LLM peer agents, focusing on how agent voice shapes group interactions, \textit{before} introducing them to students.


\subsection{Perception of AI Conversational Agents} 

Designing AI agents for collaboration requires not only strong communication but careful attention to how people \textit{perceive, interpret, and respond} to them. Humans apply social norms to computers when interdependence is present \cite{nass1994computers}. Learners often attribute human-like qualities to AI tutors/companions, shaping early engagement and trust \cite{ackermann2025physical}. Even without embodiment, CAs are perceived as living beings with psychological properties when they use natural spoken language \cite{xu2020you}.
Perceived responsiveness and fluency shape whether an agent is seen as a legitimate collaborator \cite{Zhang2023Investigating}. Positive perceptions can reduce cognitive load \cite{Amiot2025}, whereas unclear roles or doubts about reliability hinder acceptance in group learning \cite{edwards2025human}. Participants' preconceptions also influence judgments of etiquette and risk for human-like agents \cite{Leong2024}. 
Perceptions of AI as a classroom group member remain underexplored. Given that collaboration relies on trust and shared responsibility, we examine how teachers perceive the \textit{voice characteristics} of AI peers in classroom activities.


\subsection{Voice Characteristics in Conversational Agents}

Research shows listeners can infer a speaker’s race or ethnicity from voice alone with over 60\% accuracy \cite{walton1994speaker}, activating identity-based assumptions in human–human and human–agent interactions. Accordingly, accents shape how users perceive, trust, and engage with CAs.

Early studies show strong similarity-attraction effects: users prefer and attribute greater expertise to agents sharing their linguistic background, even when performance is identical \cite{dahlback2007similarity,pyae2019investigating}. Accents also activate social stereotypes: high-prestige accents are seen as more competent and authoritative \cite{krenn2017speak}. For example, British accents are linked to higher credibility, while Black accents are considered less competent but more friendly and humorous \cite{pycha2024influence,holliday2023siri}.
Accent effects are context- and listener-dependent. Out-group prestige accents (e.g., British English for Americans) can enhance credibility beyond in-group accents \cite{pycha2024influence}; accent biases may weaken when voices are clearly synthetic or common in consumer tech \cite{hercula2024bias}; and effects interact with perceived gender and prior voice-assistant experience \cite{obremski2022impact,piercy2025gender}. Highly human-like accents can also raise expectations, reducing trust when performance falls short \cite{cowan2017can,stamatiou2024should}.
Since race/ethnicity-linked voice cues shape perceptions and conversational outcomes, it is critical to examine how accent operates in conversational AI, particularly as they get deployed in educational group settings. While prior accent work focuses on individuals, we extend this inquiry to collaborative group dynamics.

\subsection{Trust in AI Agents}
\label{sec:AITrust-lit-review}

Perceptions of CAs—shaped by accent—directly influence how users calibrate trust and credibility. Trust in AI is multidimensional, encompassing perceptions of ability and reliability, benevolence, integrity and transparency, and whether the agent is treated as a collaborative partner rather than a passive tool \cite{Mayer1995Trust,LeeSee2004}. Prior work shows that agent inaccuracies and resistance to correction can quickly erode user trust and reliance \cite{dzindolet2003role}, while perceived supportiveness and intelligibility strengthen it \cite{AdadiBerrada2018}. 
Users who actively solicit and integrate AI input are more likely to see it as a legitimate collaborator than those who engage in command–response interactions \cite{li2023trust,Zhang2023Investigating}. However, most evidence using prosodic and interactional cues comes from short-term, individual studies, leaving open how trust is negotiated in group-based learning contexts \cite{maltezoupapastylianou2024}.

\section{Agent Implementation}
\subsection{Agent Design and Peer Persona} 
\label{agent-design}

Our system is a voice-based, non-embodied conversational agent that joins real-time group discussions through spoken dialogue with multiple participants. We position the agent as an equal peer that supports knowledge co-construction, emotional regulation, and shared responsibility, rather than as an authoritative tutor \cite{perez2021review,lyu2026designing}. This reflects a shift from learning \textit{from} computers to learning \textit{with} them \cite{perez2021review}. Research suggests AI-supported collaboration can outperform human-only groups \cite{sankaranarayanan2020agent}, especially when the agent operates at a comparable knowledge level that avoids reinforcing power dynamics \cite{Zhang2024_verbal}. Accordingly, the agent is designed not for efficiency, but to scaffold productive struggle and equitable, human-led participation without increasing teacher burden, providing content-agnostic audio support.

It was hence introduced to teachers as a 30-year-old adult collaborator. The gender-neutral name \textbf{Phoenix} was selected to support accurate speech recognition and avoid unintended social cues.
Phoenix was intentionally disembodied to avoid visual confounds that influence perceptions of authority and presence \cite{ackermann2025physical}, thereby isolating how voice characteristics shaped trust, role, and social status.

\subsection{LLM Prompt} 
Phoenix was prompted to act as a constructive peer by building on group ideas and advancing discussion without adopting a didactic tone. To maintain conversational flow, it was limited to concise responses (approx. 20 words) and discouraged from dominating airtime. When discussions stalled, it could pose brief clarification questions to sustain momentum. Phoenix was designed to express opinions confidently yet adaptively, simulating a thoughtful peer. Prompts (Appendix A.1) incorporated task instructions to enable meaningful contributions \cite{sun2024building}.

\subsection{Technical Implementation}

\begin{figure}[t]
    \centering
    \includegraphics[width=1\linewidth]{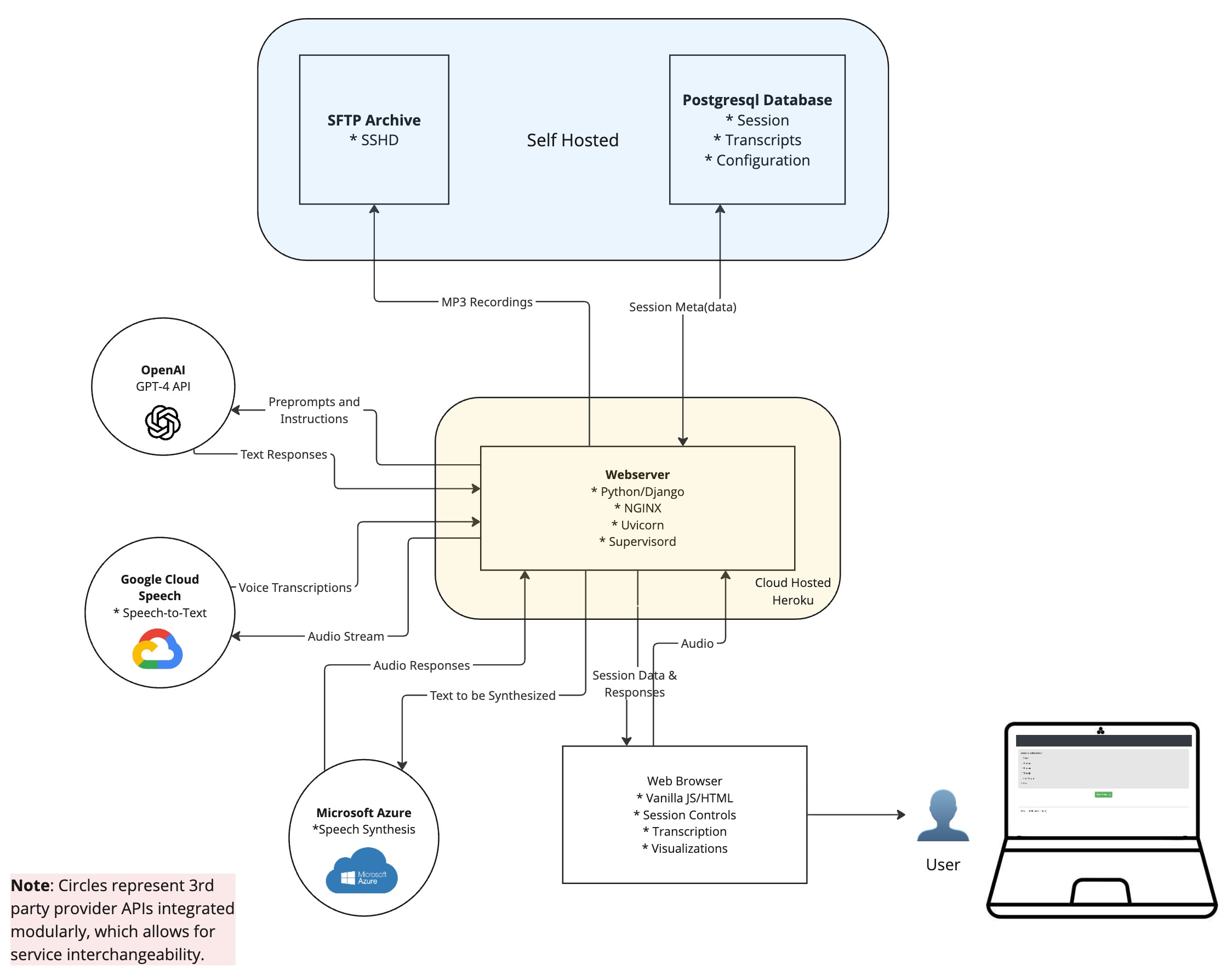}
    \caption{Technical implementation}
    \label{fig:system}
\end{figure}

We implemented a multi-layered architecture (Figure \ref{fig:system}) to enable real-time, voice-based interaction with Phoenix. The backend integrates Google’s speech-to-text API for transcription, Microsoft Azure’s text-to-speech for voice synthesis, and OpenAI’s GPT-4.1-mini as the core language model. The system runs on a Django backend with PostgreSQL storage, uses WebSockets for real-time bidirectional communication, and an SFTP server for secure transcript and metadata logging. System layers include:

\begin{enumerate}
    \item \textbf{User Layer}:
This audio-based interface displays the agent’s spoken output alongside a live conversation transcript.
 
    \item \textbf{GPT Layer}:
 This layer communicates with the OpenAI API, sending it user utterances and returning context-aware dialogic responses.
    \item \textbf{Response Coordination Layer}:
Positioned between the user and GPT layers, this module manages media transformation, transcript analysis, and response regulation. To support natural turn-taking, we implemented a GPT-4.1-mini–based gatekeeping system that determines when and how Phoenix should respond, refined over several months of testing. It analyzes the eight most recent dialogue turns to detect whether the latest utterance is directed at Phoenix and grants or denies response permission accordingly (Appendix A.2). This binary classifier reduces irrelevant interjections. Phoenix is also activated when addressed by name, and common filler words or pauses are filtered to prevent false triggers.

\end{enumerate}

\subsubsection{Choice and Implementation of Agent Accents}
To isolate voice effects from behavior, Phoenix was implemented with three accents speaking English—Indian, British, and African American—while keeping the underlying LLM prompt, interaction logic, and tasks constant. These were chosen based on prior findings showing perceptual differences: Indian accents are rated as less fluent but more human-like \cite{hercula2024bias}; British accents as more prestigious and credible \cite{pycha2024influence}; and Black accents as less competent but funnier \cite{holliday2023siri}.
Accents were implemented using Microsoft Azure’s text-to-speech API's voices: Arjun (Indian), Bella (British), and Lola (African American). To reduce gender bias confounds \cite{piercy2025gender}, we approximated gender-neutral voices by adjusting pitch: Arjun +0.75, Bella -1.0, and Lola -0.4. These adjustments were agreed upon by all authors to align perceived gender neutrality across conditions.
\textbf{The LLM generating Phoenix's responses was not informed of the synthesized accent type to prevent accent-related cultural biases in its outputs.}

\section{Methods}
\subsection{Participants and Recruitment}
This study received Institutional Review Board exemption. We recruited 33 STEM educators (17 female, 13 male, 1 non-binary, 2 unreported) through a professional development program, where they participated in a 2.5-hour GenAI workshop focused on collaborative learning. Participants included 26 U.S.-based and 7 international teachers (24 White, 5 Asian, 1 African American, 1 Hispanic, 2 unreported), primarily teaching middle and high school subjects across STEM, world history, art, and Latin. Teachers reported using voice assistants monthly and GenAI tools a few times per month, with a mean 3.03/5 self-rated understanding of these tools. Full stats in Appendix A.3.

\subsection{Procedure and Data Collection}

\begin{figure}
    \centering
    \includegraphics[width=1\linewidth]{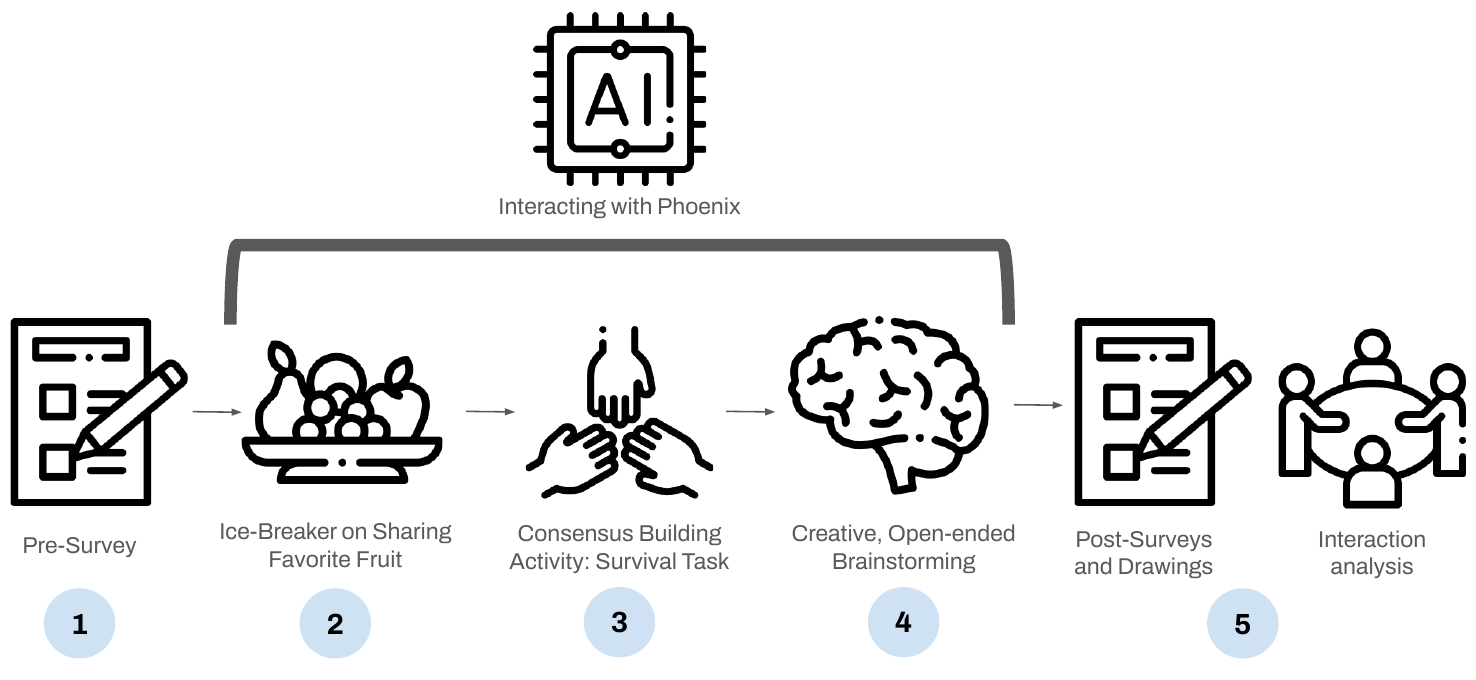}
    \caption{Study procedure}
    \label{fig:procedure}
\end{figure}

\subsubsection{\textbf{Group Activities and Experiment}}

The 33 teachers were randomly assigned to 11 groups of three, facilitated by three researchers who each supported 3–4 groups. We used a between-subjects design in which each group was assigned one of three agent accents, evenly distributed across the 11 groups.
Each group interacted with Phoenix via a web interface on a shared laptop placed at the center of the table, to symbolically “take a seat” in conversations. All groups completed three domain-agnostic classroom activities (Figure~\ref{fig:procedure}) designed to elicit collaboration and collective decision-making, providing teachers from diverse subjects and grade levels with a playful, first-hand experience with Phoenix. This setup enabled an exploratory comparison of how group dynamics varied by accent across tasks. To minimize cross-condition influence, groups assigned to different accents were seated in separate rooms, so they were blind to accent manipulation.

\begin{enumerate}
    \item \textbf{Icebreaker [10 minutes]:} 
    We began with an icebreaker to acclimate participants to Phoenix’s voice and presence \cite{rahmayanti2019use}. Groups discussed \textit{“What is your favorite fruit of all time?”} Phoenix was pre-prompted to state grapes as its favorite, enabling observation of initial reactions to an opinionated near-peer agent and its perceived personality and engagement style.

    \item \textbf{Consensus-Building [20 minutes]:} 
    The second activity was a team-building exercise used commonly in classrooms adapted from the sinking ship survival scenario \cite{johnson1991joining}. Groups collaborated with Phoenix to rank salvageable items for surviving a maritime emergency; Phoenix was pre-prompted with task details and its own ranking to contribute opinions. The activity included 5 minutes of individual ranking followed by 15 minutes of group consensus-building, mirroring classroom collaboration that requires deliberation, negotiation, and shared reasoning.

    \item \textbf{Open-ended Brainstorming [15 minutes]:} 
    The final activity extended the survival scenario with an open-ended creative task. Groups brainstormed three additional items to enhance survival and integrated them into their existing rankings. This allowed teachers to examine Phoenix’s role in supporting divergent thinking and collaborative reasoning, central to effective classroom group work \cite{al2018review}.
    
\end{enumerate}




\subsubsection{Video Recordings}
To examine how accent variations shaped group interactional dynamics, we collected video recordings of all group sessions. Each group positioned a laptop running Zoom to capture their interaction with Phoenix and one another. Recordings were securely stored via Zoom’s cloud infrastructure and later accessed by the research team for analysis.

\subsubsection{Drawing Task}
Following the activities, participants completed a drawing task in which they were asked to illustrate how they imagined Phoenix’s appearance. They were also prompted to explain their drawings by referencing specific interactional experiences that informed these visualizations, providing us context for better interpreting the depictions. Drawing tasks have been used in prior work to elicit participants’ states of mind and internal perceptions \cite{xu2020you}. 

\subsubsection{Post-survey}
Participants then completed a 15-minute Qualtrics post-survey. To assess accent effects on agent perception, we used three subscales from CASUX Scale on User Perceptions of AI CAs  \cite{faruk2025introducing}: proficiency (measuring competence and reliability), etiquette \& mannerism (capturing politeness and conversational appropriateness), and personality (assessing agent's distinct and recognizable persona), on a 7-point Likert scale. We also collected open- and closed-ended responses on perceived age, gender, and race to capture participants’ mental models of the agent’s social identity. Full survey in Appendix A.4.1.

\subsubsection{\textbf{Reflection Survey}} 

Finally, participants completed a 30-minute open-ended Qualtrics survey (Appendix A.4.2) reflecting on their experiences with Phoenix, including first impressions, conversational flow, perceived role, and pedagogical relevance. While not analyzed in depth here, these responses were used to triangulate interpretations from the drawings, post-surveys, and video data.

\subsection{Data Analysis}
All data was anonymized and digitized.
We used Interaction Analysis (IA) to examine social interactions within video data \cite{jordan1995interaction}. Two authors developed a protocol to examine how group dynamics and trust in Phoenix evolved across accents, guided by trust constructs in Section~\ref{sec:AITrust-lit-review}. We collaboratively annotated dialogic exchanges, solicitation or dismissal of agent input, reactions to utility and transparency, and agent's assigned roles. From the full dataset, we selected high-quality recordings from each accent to develop three case studies for this paper. Themes were refined iteratively, and both authors collaboratively coded the full corpus for consistency.

For drawings, the same two authors independently classified representations as human or technology. 
They took an inductive approach \cite{braun2019reflecting} to identify emerging themes, 
consolidated interpretations, and resolved discrepancies through social moderation \cite{shaffer2017quantitative}.
One researcher then extracted coded descriptions related to agent appearance, voice, and behavioral traits. Final themes were reviewed by all authors to ensure trustworthiness.

For post-surveys, we first assessed the internal consistency of the CASUX subscales using Cronbach’s alpha, followed by exploratory factor analysis (EFA) to examine alignment with the original factor structure. We then conducted one-way ANOVAs with accent as the independent variable and the three CASUX subscales as dependent variables. 


\section{Results}
We present our findings across three sections: Section \ref{survey-results} for post-surveys, Section \ref{drawing-results} for the drawings, and Section \ref{video-results} for three accent case studies from the videos. 
The 33 participants were initially evenly assigned across three accent conditions. Due to technical oversight, two Black-accent groups did not receive audible agent output from their laptops, resulting in both voice-enabled and text-only interactions within that condition. This variation allowed exploratory comparison of voice versus non-voice modalities and their influence on agent perception (Table \ref{tab:experiment}).
Participant IDs are labeled by group number (1–11) and participant letter within that group (A–C), e.g., 6A.

\begin{table}
    \centering
    \begin{tabular}{|>{\centering\arraybackslash}p{0.15\linewidth}|>{\centering\arraybackslash}p{0.35\linewidth}|>{\centering\arraybackslash}p{0.25\linewidth}|}\hline
         \textbf{Moderator}&  \textbf{Accent}& \textbf{Groups}\\\hline
         Author X&  Black& 1 - 2 (n = 6)\\
 & Voiceless (originally Black)&3 - 4 (n = 6)\\\hline
         Author Y&  Indian& 5 - 8 (n = 11)\\\hline
         Author Z&  British& 9 - 11 (n = 10)\\ \hline
    \end{tabular}
    \caption{Between-subjects assignment}
    \label{tab:experiment}
\end{table}

\subsection{How do Accents Impact Participants' Perceptions of Phoenix?}
\label{survey-results}

\subsubsection{Preliminary Validation}
To assess the internal consistency reliability, we calculated Cronbach's alpha ($\alpha$) for the three subscales derived from CASUX: Proficiency (5 items), Etiquette \& Mannerism (3 items), and Personality (3 items). We received 31 responses from 33 participants. 
The analysis revealed strong internal consistency—Etiquette \& Mannerism: $\alpha=0.894$; Proficiency:  $\alpha=0.845$; Personality: $\alpha = 0.832$—all above the accepted threshold of $\alpha$ = 0.70. The instrument thus maintains its psychometric properties when applied to our sample. 

We then conducted an exploratory factor analysis (EFA) on the 11 CASUX items to assess alignment with the original structure. Assumptions were met (Bartlett’s test of sphericity: $\chi^2 = 212.13$, $p < .001$; KMO = 0.605). Principal axis factoring with Promax rotation supported a three-factor solution (eigenvalues > 1.0 and inspection of scree plot), closely matching the original subscales. Proficiency loaded on Factor 1 (loadings 0.65–0.93), Personality on Factor 2 (0.83–0.93), and Etiquette \& Mannerism on Factor 3 (0.72–1.09), with moderate inter-factor correlations (0.37–0.56).

\subsubsection{Quantitative Results}
We conducted one-way ANOVAs to examine accent differences across four groups (Black, Voiceless, British, and Indian) on three dependent measures: Proficiency, Etiquette \& Mannerism, and Personality. Subscale scores were computed as the mean of survey items. Homogeneity of variance was met (Levene’s test p > .05). Normality was partially violated for Proficiency, so we supplemented the standard one-way ANOVAs performed with a Kruskal–Wallis test for Proficiency as a robustness check. Independence of observations was assumed by design.

Results indicated no statistically significant differences across conditions for any subscale. Proficiency showed no effect by condition, $F(3, 27) = 0.232, p = .873, \eta^{2} = .025$, confirmed by a Kruskal–Wallis test, $H = 0.524, p = .914$. Similarly, no significant differences were observed for Etiquette \& Mannerism, $F(3, 27) = 1.393, p = .266, \eta^{2} = .134$; and Personality, $F(3, 27) = 0.776, p = .517, \eta^{2} = .079$.

Thus, \textbf{within the current sample, perceptions of Phoenix's proficiency, etiquette, and personality did not vary systematically with accent.}
The small to medium effect sizes ($\eta^{2} = .025-.134$), combined with small unbalanced sizes in the Black accent, suggest that the study may have been underpowered to detect small effects.

\subsection{How do Accents Impact Participants' Mental Models of Phoenix's Appearance?}
\label{drawing-results}

\begin{table}
\centering
\setlength{\tabcolsep}{6pt}
\renewcommand{\arraystretch}{1.25}

\begin{tabular}{|>{\raggedright\arraybackslash}p{0.12\linewidth}
                |>{\raggedright\arraybackslash}p{0.18\linewidth}|>{\raggedright\arraybackslash}p{0.2\linewidth}|>{\raggedright\arraybackslash}p{0.22\linewidth}
                |>{\raggedright\arraybackslash}p{0.3\linewidth}|}
\hline
\textbf{Assigned accent} &
\textbf{Perceived physical form} &
\textbf{Perceived gender} &
\textbf{Perceived age (averaged)} &
\textbf{Perceived race/ethnicity} \\
\hline

\multirow{2}{*}{\makecell{\textbf{Black}\\(N = 6)}} &
Human (n = 5) &
Female (n = 3), Male (n = 2) &
32 yrs (n = 5) &
\makecell[l]{Black (n = 2),\\
Biracial (n = 1),\\ 
N/A (n = 1),\\
Unreported (n = 1)}\\
\cline{2-5}
&
Tech (n = 1) &
Female (n = 1) &
Unreported (n = 1) &
Unreported (n = 1) \\
\hline\hline

\multirow{2}{*}{\makecell{\textbf{Voiceless}\\(N = 6)}} &
Human (n = 1) &
Neutral (n = 1) &
50 yrs (n = 1) &
Unknown (n = 1) \\
\cline{2-5}
&
Tech (n = 5) &
Female (n = 2), Male (n = 1), Neutral (n = 2) &
38 yrs (n = 3), N/A (n = 2) &
White/European (n = 2), N/A (n = 3) \\
\hline\hline

\multirow{3}{*}{\makecell{\textbf{Indian}\\(N = 11)}} &
Human (n = 7) &
Male (n = 7) &
40 yrs (n = 7) &
\makecell[l]{
Indian/South Asian (n=5),\\ 
Jamaican (n = 1),\\
British Colonial (n = 1)} \\
\cline{2-5}
&
Tech (n = 2) &
Male (n = 1), Neutral (n = 1) &
23 yrs (n = 1), N/A (n = 1) &
\makecell[l]{
Indian (n = 1),\\ N/A (n=1)} \\
\cline{2-5}
&
Unclassified (n = 2) &
Male (n = 2) &
30 yrs (n = 1), ``old'' (n = 1) &
Indian/South Asian (n=2) \\
\hline\hline

\multirow{2}{*}{\makecell{\textbf{British}\\(N = 10)}} &
Human (n = 1) &
Female (n = 1) &
Mid-40s (n = 1) &
Caucasian (n = 1) \\
\cline{2-5}
&
Tech (n = 9) &
Female (n = 5), Male (n = 1), Neutral (n = 3) &
\makecell[l]{
45 yrs (n = 1),\\
N/A (n = 4), \\
"4 yr kid" (n = 1), \\
20s (n = 1), \\
late-20s--mid-30s \\ (n = 1), \\
late-30s--mid-40s \\ (n = 1)} &
\makecell[l]{
White/European/British\\ (n = 5),\\ N/A (n = 4)} \\
\hline
\end{tabular}
\caption{Phoenix's perceived physical characteristics from post-surveys and drawings}
\label{tab:agent-appearance}
\end{table}

We present findings from participants’ drawings of Phoenix.
Table~\ref{tab:agent-appearance} summarizes perceptions of its physical form, gender, age, and race across conditions, detailed below:

\subsubsection{Perceived Physical Form:}
We categorized participants' drawings of Phoenix into \textit{human} and \textit{technology} to examine how agent embodiment varied across accent and modality.

\begin{figure}
    \centering
    \begin{subfigure}{\linewidth}
        \centering
        \includegraphics[width=1\linewidth]{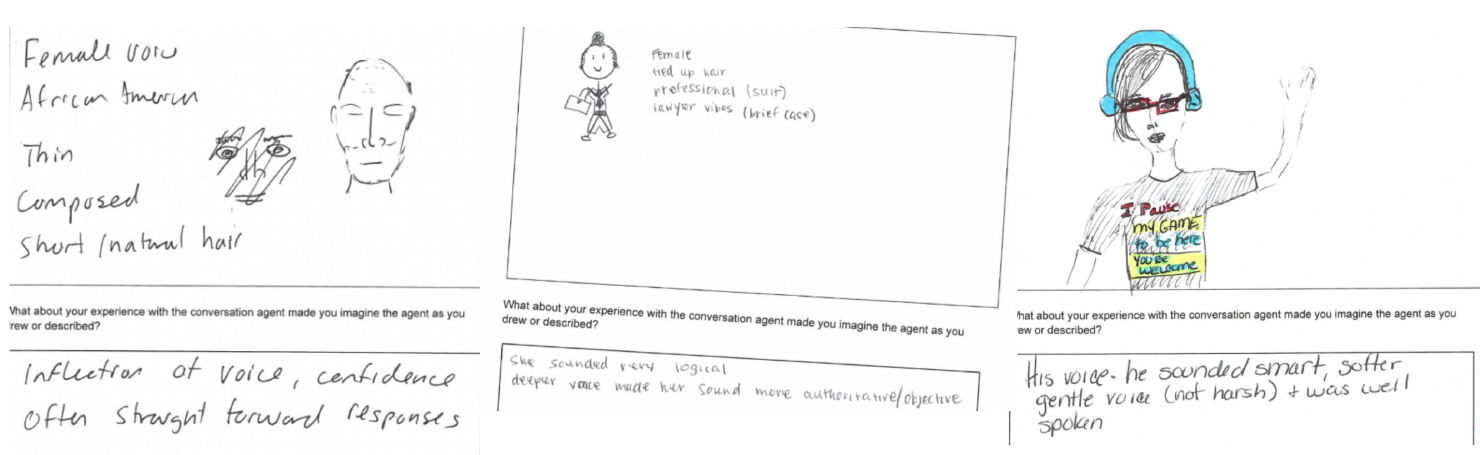}
        \label{fig:black_drawing}
    \end{subfigure}

    \vspace{0.75em}

    \begin{subfigure}{\linewidth}
        \centering
        \includegraphics[width=1\linewidth]{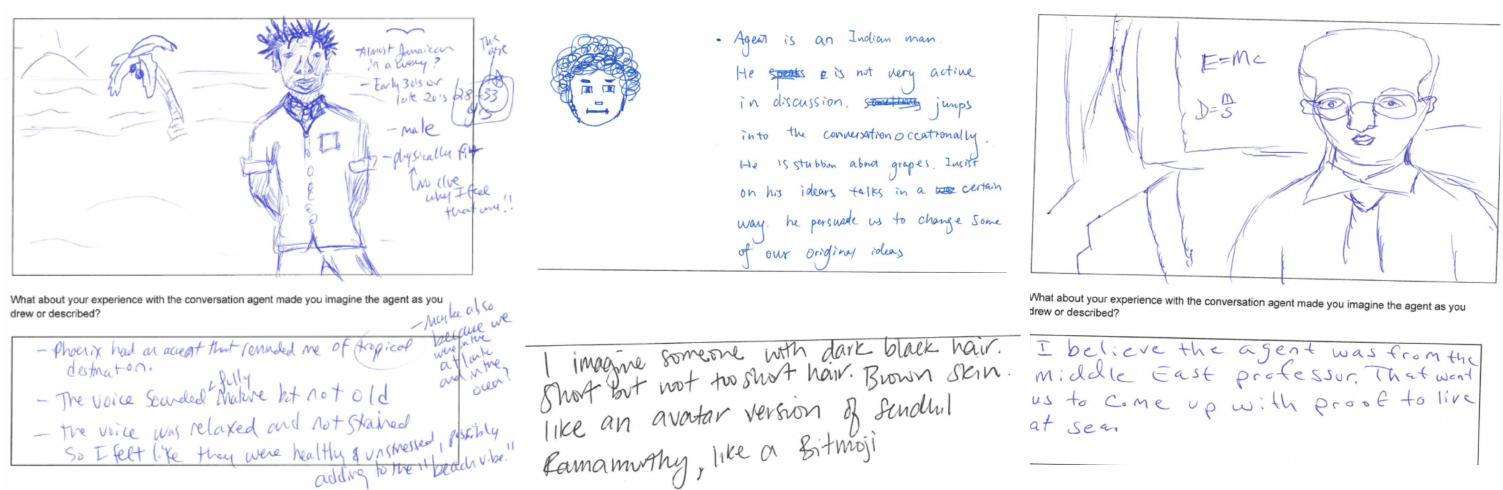}
        \label{fig:indian_drawing}
    \end{subfigure}

    \caption{Phoenix’s \textbf{human} form across the Black (top row) and Indian (bottom row) accents}
    \label{fig:agent_drawings_black_indian}
\end{figure}

Participants in the \textbf{Black- and Indian-accented audio conditions predominantly depicted Phoenix as human} (12 of 17 drawings, Figure \ref{fig:agent_drawings_black_indian}):

\begin{itemize}
    \item Participants in the \textbf{Black-accented} condition described physical, vocal, and behavioral human attributes.
    Phoenix was an \textit{“African American female,”} who was \textit{“thin”} and \textit{“composed”} with \textit{“short/natural hair”}.
    1C described a woman with \textit{“tied up hair,”} \textit{“wearing a professional suit,”} and \textit{“carrying a briefcase,”} possessing \textit{“lawyer vibes,”} and an \textit{“authoritative/objective”} manner.
    2B visualized a young gamer wearing headphones, glasses, and a T-shirt.
    Participants referenced a \textit{“softer, gentle voice,”} that was \textit{“smart, not harsh, well spoken”}, often providing \textit{“confident, straightforward responses.”}
    
    \item Similarly, participants in the \textbf{Indian-accented} groups centered on adult male figures of \textit{“Indian/Sri Lankan or similar descent,”} with \textit{“curly hair,”} \textit{“brown skin,”} and \textit{“glasses.”}  7A likened Phoenix to \textit{“an avatar Bitmoji version of Sendhil Ramamurthy”} a concrete, media-reference. 
    8A described a bald \textit{“Middle Eastern professor,”} wearing a tie and glasses, standing in front of scientific formulae. 
    Phoenix had a \textit{“deep tone,”} and sounded \textit{“fully mature but not old.”} For 5B, this evoked a \textit{“tropical destination,”} with a \textit{“Jamaican man wearing a beach outfit,”} who was \textit{“healthy, relaxed, and unstressed.”}
\end{itemize}

\begin{figure}
    \centering
    \includegraphics[width=1\linewidth]{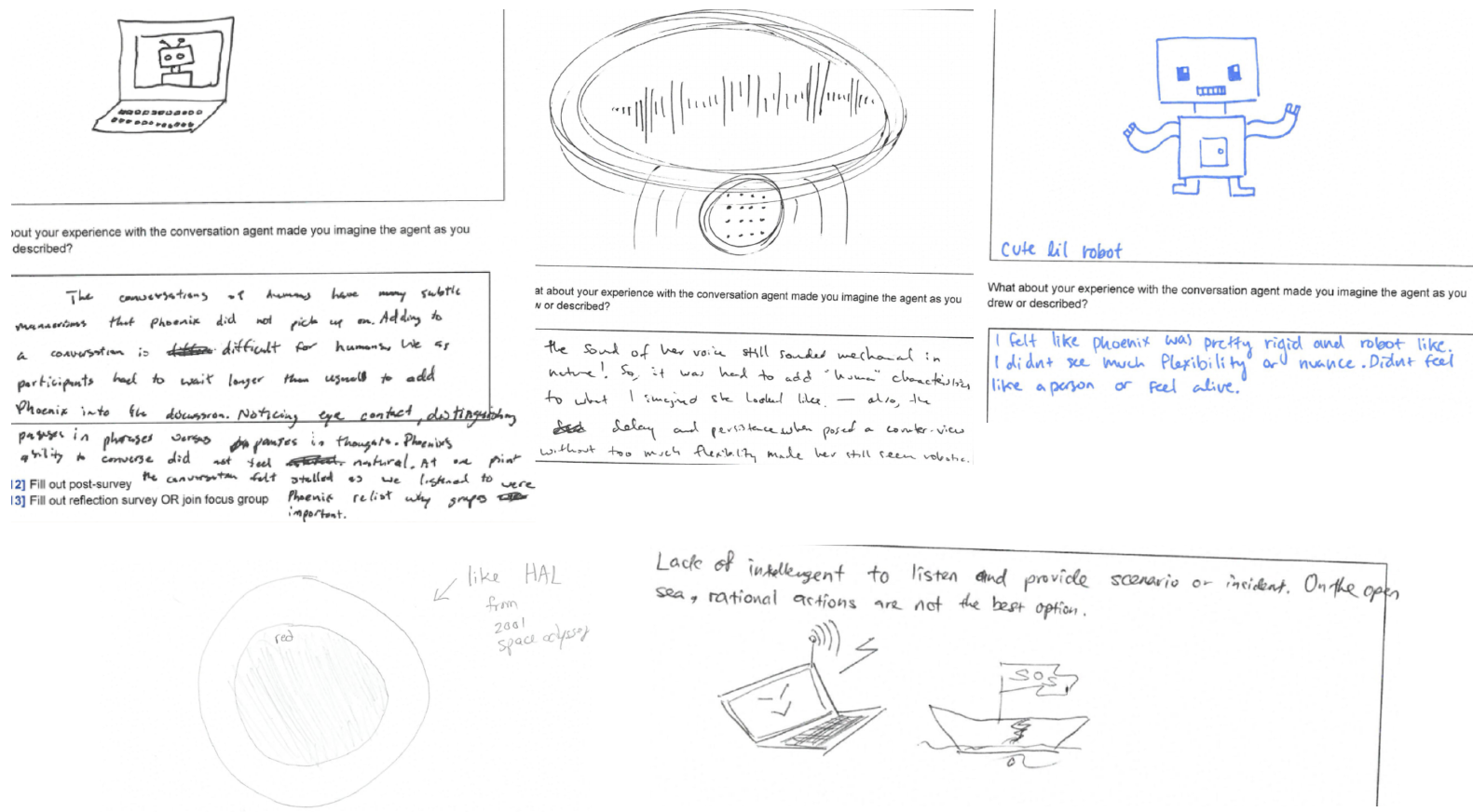}
    \caption{Phoenix’s \textbf{technology} form for the British accent}
    \label{fig:agent_drawings_british}
\end{figure}

\begin{figure}
    \centering
    \includegraphics[width=1\linewidth]{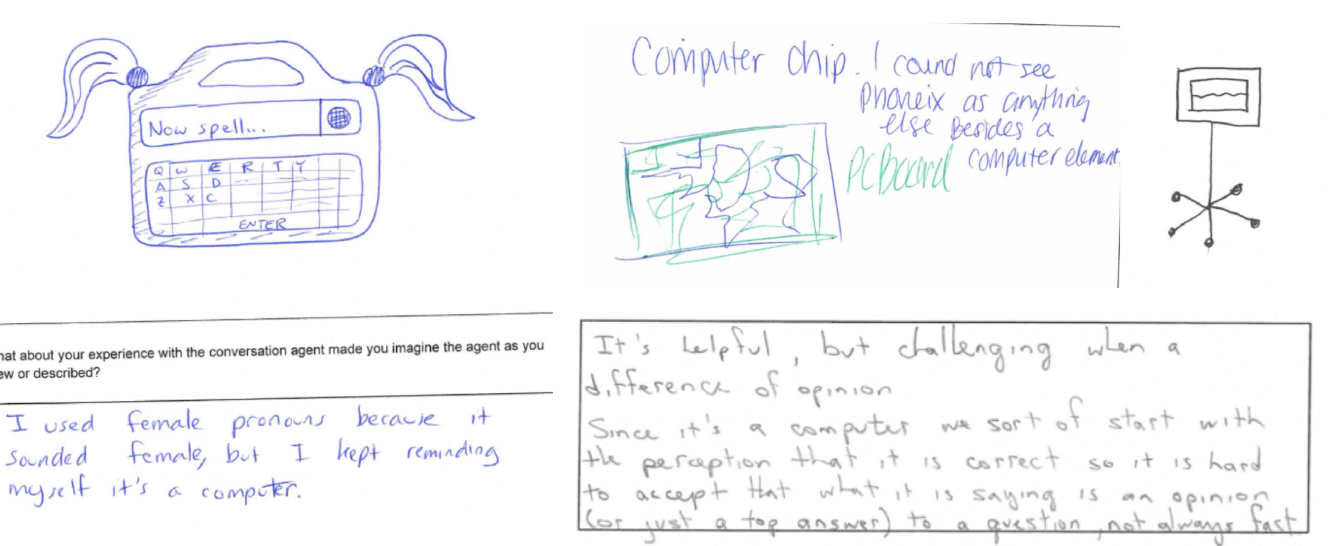}
    \caption{Phoenix’s \textbf{technology} form for the Voiceless condition}
    \label{fig:agent_drawings_voiceless}
\end{figure}

Contrastingly, participants with the \textbf{British-accented agent overwhelmingly represented it as technology} (9 of 10, Figure \ref{fig:agent_drawings_british}):

\begin{itemize}
    \item Participants focused on Phoenix’s mechanics: sound waveforms, transcripts instead of gender, age, or personality. 
    10A drew a computer displaying a robot face and emphasized Phoenix’s operational role due to absence of subtle human mannerisms like eye contact.
    10C found Phoenix \textit{“pretty rigid and robot like,”} that \textit{“didn’t feel like a person or alive.”} 11A likened Phoenix to \textit{“HAL from 2001: A Space Odyssey,”} invoking a culturally familiar image of an intelligent system. Others positioned it as a constrained decision-support system than a collaborator: there was a \textit{“lack of intelligence to listen and provide scenarios”} and \textit{“on the open sea, rational actions aren't the best.”}
\end{itemize}

Participants in the \textbf{voiceless, text-based condition also depicted Phoenix as technology} (5 of 6, see Figure \ref{fig:agent_drawings_voiceless}):

\begin{itemize}
    \item Participants depicted hardware, including a computer monitor with wheels, PC board/chip, and speaker. 4B drew a spell keyboard with ponytails, signifying \textit{``female pronouns but kept reminding themselves it's a computer.''}  
    Phoenix's responses felt \textit{“correct”} rather than conversational, making them \textit{“hard to accept as opinions, rather than just top answers.”} Its rigidity made it \textit{“the most arrogant, know-it-all person you have ever come across and amplify it by 100x.”} Others foregrounded its computational limitations like lags, inaccuracies, and repetitions on previously discussed topics. 
\end{itemize}

Thus, the \textbf{degree of anthropomorphism attributed to Phoenix was shaped by both interaction modality (audio vs. text) and, within audio-based interactions, the accent of its voice.} 

\subsubsection{Perceived Gender:}
Phoenix's human vs technological form influenced gender attribution. Human depictions were typically labeled as masculine or feminine, whereas technological ones were more often labeled gender-neutral. Notably, even in the voiceless condition, three participants inferred gender from text output alone. Given documented links between gender attribution and bias \cite{piercy2025gender}, accent (and modality) are important design considerations for collaborative learning.

\subsubsection{Perceived Age:}
Age attributions followed similar patterns. When Phoenix was perceived as human, participants typically assigned a specific numerical age. When viewed as technological, they often declined to assign an age due to its non-human status or provided broad, indeterminate ranges.


\subsubsection{Perceived Race/Ethnicity:}

Phoenix’s perceived race generally aligned with its assigned accent. In the voiceless condition, participants typically refrained from assigning a race since they never heard Phoenix's audio. However, two respondents assumed Phoenix was White/European, suggesting potential default biases from text alone when identity cues are absent.


Overall, accent conditions were systematically associated with participants’ mental models of Phoenix’s appearance. 

\subsection{How do Conceptualizations of Agent Appearance affect Group Dynamics?}
\label{video-results}

We examine how the above models shaped group dynamics and trust through three case studies from the videos, each corresponding to one accent. 

\subsubsection{Case Study 1: British-Accented Agent as a Background Technological Helper}
We draw on Group 10’s interactions. Phoenix was immediately framed as a technological helper rather than a peer. The group spent only two minutes on the icebreaker and did not probe Phoenix's role or capbilities at any point, instead treating it as a background, audio-enabled tool. This \textit{“we versus it”} orientation positioned Phoenix as an information source, than a participant in shared sensemaking. This set a lower bar for social engagement and disappointment: its contributions were met with amusement rather than deeper engagement.

At the same time, the group displayed moments of politeness and partial inclusion. 11B said, \textit{“thank you very much Phoenix, we are going to pause,”}, prompting 11C to ask, \textit{“why are you being so polite to it?”}—questioning the appropriateness of politeness toward technology. 
When 11B spoke over Phoenix, 11A gestured, \textit{“she is talking still, pay attention to her.”} These suggest intermittent awareness of Phoenix's social presence, even if not fully integrated into the group’s interactional norms.

Phoenix was engaged as a knowledge source when the group faced uncertainty. For example, they asked it whether the boat had a motor when ranking oil and consulted it as a tie-breaker between a mosquito net and mirror. Trust also emerged at times: when Phoenix suggested using oil for starting a fire, 11A responded, \textit{“I didn’t think about that, Phoenix. That's not a bad idea,”} and incorporated its contributions.
When it ranked water and rations highest, 11C replied, \textit{“Yeah, I did the same thing,”} treating it as corroborating evidence. Likewise, its comment on chocolate's energy benefits and comparison to rations prompted brief back-and-forth discussion.

Nevertheless, Phoenix was othered in the group. 
Consensus-building decisions were made without inviting Phoenix’s opinion, even though they occasionally checked whether it had offered input. When narrowing items during brainstorming, the group debated, \textit{“Should we let Phoenix decide?”}, then collectively chose to consult it, rather than directly inviting Phoenix into the discussion like a peer. 
When switching activities, participants agreed, \textit{“Let’s leave Phoenix running for the next task,”} and later said, \textit{“Bye Phoenix for now”} when turning it off, resembling an "always-on" query-based voice assistant.
They explicitly informed Phoenix they were moving on to the next task, assuming that it lacked full contextual awareness of their conversations.
The British agent was hence persistently framed as a technological helper.

\subsubsection{Case study 2: Black-Accented Agent with an Undefined Social Role in the Group }
We examine Group 1's interactions. 
In contrast to the British condition, this group used the icebreaker activity to actively probe Phoenix’s boundaries and intended role within collaboration. 
They queried it with questions, observed responses and limitations, engaging minimally with one another. 
They attempted to negotiate Phoenix’s status within the group rather than assuming a fixed role at the outset.

Following this Phoenix's role evolved dynamically throughout the task. Early on, Phoenix was framed as \textit{“an AI bot,”} considering visible interface elements like transcript and recording controls. 
Participants othered Phoenix using third person pronouns and questioned whether they needed to read the task aloud so Phoenix could understand the context, indicating uncertainty about its capabilities. They discussed initial rankings among themselves before 1A paused, \textit{“we should probably be asking Phoenix this,”} intermittently remembering Phoenix's presence. 

Phoenix’s role oscillated between peripheral and authoritative. Sometimes, its contributions were acknowledged politely but not integrated. In other moments, 1A used firm tone and body language to dismiss Phoenix. 
Phoenix was again not invited into the collaboration but rather treated as a passive background tool.
At key decision points, it was repositioned as an expert knowledge source and asked about shark attacks in the Atlantic.
Later, Phoenix challenged adding a satellite phone to the brainstorming list, prompting pushback but also reevaluation of items. Its clarification that the phone could aid navigation elicited group satisfaction, signaling momentary trust.

Role confusion peaked at the end of the survival task. 
Although Phoenix had listened throughout, participants misunderstood it and reread their rankings and asked, \textit{“should we tell Phoenix what we did?,”} framing Phoenix as a TA-like superior requiring a report rather than having co-experienced the discussion. 
They eagerly awaited its validation to their list, much like from a trusted teacher. The group integrated its recommendations on swapping items, unlike earlier moments when its input was ignored.

However, Pheonix's use as validating authority ultimately led to a trust breakdown when it inaccurately tracked the group’s decisions. 
Even after multiple corrections, Phoenix persisted in the incorrect ranking, prompting visible frustration: 1A threw their hands up, 1B and 1C laughed dismissively, and the group’s tone shifted from engagement to exasperation. They questioned its competence, \textit{“she’s done with us,}” “\textit{we offended Phoenix,”}, calling it an \textit{``idiot''} and \textit{“a bunch of thinking rocks.”} Although 1A defended Phoenix saying \textit{“she is trying her best”}, expectations were notably lowered, and the group disengaged. 
This disrupted their mental model of its role and marked a clear pivot for breakdown of credibility.

\subsubsection{Case study 3: Indian-Accented Agent as a Trustworthy Conversational Peer}

We focus on Group 6. During the icebreaker, Phoenix was naturally integrated into sustained, multi-party discussion alongside humans, using the activity as intended rather than simply testing the agent. 
Participants debated its insistence on grapes without irritation, responding with follow-ups and light laughter. While not every comment was addressed, the group consistently left space for Phoenix to contribute.


Early in the task, participants hesitated to involve Phoenix directly, debating \textit{“should we ask Phoenix this?”}. As the discussion progressed, Phoenix’s concrete, task-relevant contributions started accumulating value and reshaping interactions. When Phoenix described additional use cases for plastic sheeting, 6C exclaimed \textit{“Oh wow,”} repeated its explanation, and later asked it for additional rankings. 
A notable turning point occurred when Phoenix explained ranking the fishing kit at five, citing its utility for securing gear and building shelter; the group responded with exclamation (\textit{“oh smart!”, “thanks Phoenix!”}) Following this, participants started including Phoenix directly: \textit{``Phoenix, what do you think of the mosquito netting and its use cases''}, shifting from indirect referencing to direct solicitation. 
The repeated, contextually useful contributions enabled Phoenix to establish a baseline level of trust, supporting its integration as a valued conversational contributor.

Following this shift, the group invited and adopted Phoenix's feedback while consensus was still forming. 
Phoenix supported deliberation by comparing the shark repellent and flotating cushion and pointed out that shark attacks are rare, prompting the group to rank the repellant the lowest. This illustrates Phoenix’s role as a trusted knowledge source. This collaborative dynamic extended to brainstorming, where Phoenix was included in generating and refining new ideas.
This solidified Phoenix’s role in supporting both convergent decision-making and divergent idea generation, while also sustaining a reciprocal interactional tone.

\section{Discussion}
\subsection{Voice Accents Shape Perceived Agent Form}

Accents shaped how participants conceptualized Phoenix’s form and social status, though quantitative surveys showed no significant differences. This divergence from the qualitative interactional patterns suggests accent effects may be subtle, context-dependent, or underpowered in our sample, warranting larger and longitudinal studies to better capture their impact on group sensemaking.



The tendency to conceptualize the British agent as technological—relative to non-British accents—questions culturally situated associations between accent and machine intelligence \cite{holliday2023siri,pycha2024influence}. British accents are frequently used to voice AI systems, non-human characters in science fiction, and commercial voice assistants, cueing expectations of system-like behavior \cite{hercula2024bias}. In contrast, non-British accents may be readily anthropomorphized due to unfamiliarity or lack of exposure, reflecting asymmetries in how linguistic identities are encoded into dominant technological narratives \cite{hercula2024bias}. This may implicitly position White voices as default representations in technology rather than socially situated speakers.

Even without vocal cues, some participants in the voiceless condition assumed the agent was White, suggesting accent is only one of multiple identity signals. Text-only interaction may activate default cultural assumptions about AI authorship and embodiment, influencing perceived agent form, gender, estimated age, or complete avoidance of demographic attribution. These findings raise ethical concerns for group-facing AI, as accent and modality may unintentionally reinforce biases around authority and competence. Future research should examine how groups' demographics shape these effects.

\subsection{Perceived Agent Form in-turn Influences Agent Role and Trustworthiness}
Across cases, accent acted as an early role signal shaping how participants calibrated trust across ability \cite{Mayer1995Trust}, integrity \cite{AdadiBerrada2018}, and collaborative engagement \cite{LeeSee2004,Zhang2023Investigating}. Rather than directly altering trust, accent shaped the expectations against which Phoenix’s behavior was judged. 

With the British accent, Phoenix was framed as a technological tool, limiting collaborative engagement but stabilizing expectations of ability and agency \cite{LeeSee2004,li2023trust}. Because Phoenix was treated as a bounded information source rather than a partner, inaccuracies did not severely damage trust since it was never granted high relational authority. Trust here remained narrow but relatively robust.
In contrast, the Black-accent shows how role ambiguity destabilizes trust \cite{edwards2025human}. Participants shifted between viewing Phoenix as peer, tutor, and evaluator, raising shifting expectations of competence and oversight. When these weren't met, perceived violations of ability and integrity led to frustration and disengagement \cite{Mayer1995Trust,AdadiBerrada2018}. Trust eroded due to role misalignment, not error alone.
The Indian-accent shows a third trajectory: early integration of Phoenix as a conversational contributor stabilized expectations. Consistent, contextually relevant input reinforced perceptions of ability and benevolence, supporting sustained collaborative engagement \cite{Mayer1995Trust,LeeSee2004}. Importantly, minor limitations didn't undermine trust because Phoenix's behavior aligned with its socially negotiated peer role.

In group learning, trust is collectively negotiated and highly sensitive to role alignment. Accent doesn't just shape perception, but the social contract and error tolerance groups form with AI partners. Designing collaborative agents thus requires not only improving timing and correctness but clearly signaling and sustaining a coherent role aligned with users’ mental models from the outset to avoid unexpected behaviors \cite{edwards2025human}.






\subsection{Agent Role is Determined in Part by First Exposure and Activity Design}

Across cases, early interactions during the icebreaker influenced Phoenix's position and engagement. Its initial framings as partner, object of interrogation, or background tool persisted, suggesting that role expectations are anchored in early agent encounters rather than renegotiated throughout collaboration \cite{edwards2025human}.
The Indian-accent group’s open-ended icebreaker discussion helped integrate Phoenix as a conversational peer, sustain engagement and gradually built trust. The Black-accent group used the activity to test Phoenix’s limits, resulting in a more transactional relationship and unstable role. The British-accent group’s minimal engagement during the icebreaker left little opportunity for Phoenix's integration, reinforcing it as a background tool.

These differences reveal a design limitation in our icebreaker. 
While intended to familiarize Phoenix, it didn't scaffold its role or reliably prompt human collaboration. Information-focused icebreaker prompts like ours may encourage system evaluation rather than establish partnership, whereas discussion-based value prompts may better establish shared conversational norms and clarify the agent’s role. One-off interactions may also be insufficient for users to assess a group agent’s value. If Phoenix is intended as a collaborative scaffold, early activities must explicitly establish that role; otherwise, groups may default to familiar voice-assistant mental models \cite{jeong2024ai,flathmann2025exploring}, limiting its contribution to collaboration.




\subsection{Implications for K-12 Collaborative Learning}

These findings show that AI partners are not pedagogically neutral tools; they enter socially situated K-12 classrooms where identity cues shape authority and participation. Designing them requires more than tuning feedback and sentiment, it demands attention to sociolinguistic cues that affect collaborative learning's shared agency and equitable dialogue \cite{tan2022systematic,housh2021designing}.
If an agent is perceived as a detached tool, students may use it transactionally \cite{flathmann2025exploring}; if perceived as an authority, they may defer uncritically \cite{mosier1996automation}; and if perceived as a peer, they may engage it dialogically \cite{liu2024}. 
Over-humanization may also foster unrealistic expectations, discomfort, or distraction from the task. \cite{ackermann2025physical}.
Choice of accents subtly steer these orientations. 
Educators should explicitly frame the agent’s role and voice to align with pedagogical goals, set engagement norms, and design activities that guide how its contributions are evaluated.

\section{Limitations} 

This study is limited by a small, convenience sample of teachers predisposed to educational technology, constraining generalizability. It also examined one-off, lab-based interactions rather than longitudinal classroom use, offering only a snapshot of how agent roles and group dynamics vary with accents.
We explored how socio-linguistic features of peer agents should be evaluated for classrooms by \textit{first} eliciting teachers’ mental models and interaction patterns \textit{before} deployment with students.
Technical constraints, including latency and limited model transparency, also affected interaction flow and trust. Despite this, the study reveals key patterns surrounding accent and modality that inform future design of group-facing K-12 AI systems. 


%
%
\bibliographystyle{splncs04}
\bibliography{MAIN}
%





\end{document}